# Centimetre-scale propagation of optical surface waves at visible wavelengths

Babak Vosoughi Lahijani,[1,2,*] Nicolas Descharmes,[1] Raphael Barbey,[1] Gael D. Osowiecki,[1] Valentin J. Wittwer,[3] Olga Razskazovskaya,[3] Thomas Südmeyer,[3] and Hans Peter Herzig[1]

*[1] Optics & Photonics Technology (OPT) Laboratory, École Polytechnique Fédérale de Lausanne (EPFL), rue de la Maladière 71b, CH-2002 Neuchâtel, Switzerland*

*[2]Currently with the Department of Photonics Engineering, DTU Fotonik, Technical University of Denmark, Building 343, DK-2800 Kgs. Lyngby, Denmark*

*[3]Laboratoire Temps-Fréquence, Université de Neuchâtel, Av. de Bellevaux 51, CH-2000 Neuchâtel, Switzerland*

**Abstract:**

Guiding and confining light at interfaces is crucial for applications involving light-matter interactions such as surface spectroscopy, nonlinear optics, and quantum information technology. While dielectric surface waves offer promising perspectives for strong light confinement at non-absorbing interfaces, their capacity has not yet been widely explored. Here, we focus on the propagation properties of optical modes supported at the free surface of a one-dimensional photonic crystal. The contributions of intrinsic and extrinsic loss mechanisms are investigated. Structures optimized to support long propagating optical surface waves are designed and fabricated. We experimentally demonstrate, for the first time, the existence of optical surface waves capable of propagating over centimeter-scale distances in the visible spectral range. Our results open new perspectives for the use of optical surface waves in integrated optics and enhanced light-matter interactions.

## 1. Introduction

In their seminal 1978 paper, Yeh, Yariv and Cho[1] reported the first experimental demonstration of the optical equivalent of an electronic surface state, that is, a photonic state whose wavefunction is distributed around the truncation plane of a semi-infinite periodic potential. This optical analogue is often referred to as a Bloch surface state and is a part of a larger group of optical surface states. The latter includes D'yakonov states[2,3] and the many different types of surface polaritons, among which, the well-known surface plasmon-polaritons[4,5].

Optical surface states possess a variety of interesting features, two of which are of particular interest for practical applications and devices. First, the distribution of the field around the free interface favours light-matter interactions. This feature can be used for sensing, spectroscopy, but also for generating efficient electro-optical effects[6–8]. Second, optical surface states are propagating and can be used to efficiently transport energy and information along the interface[9].

It is thus possible to envisage, one day, the creation of surface-state-based photonic chips and circuits. These chips could include surface state coupled light emitters[10], waveguides[11,12], resonators[13–15] and detectors[16], and benefit from their great surface sensitivity through fast, power-efficient modulators and switches[17]. This platform could work in the visible range and offer additional possibilities to silicon photonics. Thin silicon detectors deposited on the surface could be advantageously used for the detection of the surface waves. The large variety of optical materials available, including gain materials, would offer a vast ensemble of design and functionalities, while ensuring compatibility with standard semiconductor technology.

In this context, whether to maximize the interaction volume of the surface field with an analyte, or to transport energy or information, the propagation length of the surface waves is a parameter of primary importance. In their work, Yeh et al. assessed the existence of an optical surface mode in the near-infrared (1.15 µm) in GaAs/AlGaAs stacks grown by molecular beam epitaxy. The authors reported, through a rough experimental determination, an extinction coefficient smaller than $10^{-2}$ $cm^{-1}$ for the fundamental surface mode[1] making thus optical surface waves relevant candidates for on-chip optical information transportation. Over the past forty years, many studies have explored the possibilities offered by Bloch surface waves, from the near-UV to mid-infrared wavelengths[18–26], particularly for sensing applications[27–32]. Still, only very few works have discussed their propagation properties[33,34], and even fewer have made reports of experimentally measured propagation lengths[35,36]. At visible wavelengths, the longest propagation lengths achieved range in the order of a few hundred micrometers[37].

In this work, we design optical surface wave supporting structures specifically optimized for long propagation in the visible range. A discussion of the different loss mechanisms of BSWs is presented, considering both intrinsic and extrinsic factors. In this discussion, the primary roles of free radiation into the substrate and scattering as key contributions to the losses are hypothesized. An eigenmode solver-based one-dimensional model is developed to study this hypothesis. The 1D model is then used to design structures that are optimized to support a long propagating surface mode. These structures are designed with identical lattice parameters but varying numbers of lattice periods so as to reveal possible extrinsic loss mechanisms. These structures are then fabricated using state of the art ion-beam sputtering so as to ensure optimal deposition quality. Finally, we individually characterize their propagation lengths using a surface scattering loss measurement method and measure a record propagation length of 1.4 cm at visible wavelengths. This is well above one order of magnitude longer than the previous reports in this part of the spectrum.

## 2. Results

**2.1. Loss analysis of Bloch surface waves.** Bloch surface waves exist along the interface separating a one-dimensional photonic crystal from a homogeneous dielectric medium and are characterized by a mode profile that is exponentially decaying, both in the superstrate medium – in this case air – and in the one-dimensional photonic crystal structure. A typical field distribution of a Bloch surface mode can be seen in Figure 1a. In this case, the one-dimensional photonic crystal consists of a stack of N = 10 lattice periods. The lattice consists of $Ta_2O_5$ and $SiO_2$ layers with respective thicknesses of 107 nm and 158 nm. The periodicity of the stack is interrupted by an additional 34 nm thin $Ta_2O_5$ defect layer. This structure supports a TE-polarized surface mode at visible wavelengths.

The dispersion curve of the optical surface mode is computed using the CAvity Modelling FRamework (CAMFR) eigenmode solver[38]. It is shown as a red line in the dispersion diagram of Figure 1b. The mode dispersion extends from the air line to the $SiO_2$ line (indicated as black solid lines), which it finally intersects. For a given lattice, the position of the dispersion curve is mostly dictated by the parameters of the top most layer terminating the structure.

Here, three main loss mechanisms that might limit the propagation length of Bloch surface waves are envisaged: (i) Radiation into the substrate. The dispersion line of the mode is mostly located above the fused silica substrate ($SiO_2$) line. This means that energy propagating in the mode within this range of frequencies can freely couple to the substrate assuming that the parallel component of the wavevector is conserved. Bloch surface modes above the $SiO_2$ line are thus inherently radiating modes. The magnitude of this loss can be seen as the overlap of the surface mode with the radiation continuum. Increasing the number of lattice periods that physically isolate the surface mode from the substrate should thus decrease this unwanted coupling to the continuum. This loss mechanism is intrinsic to the nature of the mode. (ii) Light scattering induced by the surface roughness and layer defects. The amount of scattering is related to the quality of the fabrication and thus, is extrinsic to the surface mode. (iii) Material absorption. A large fraction of the mode is located within the periodic structure, and propagation takes place perpendicular to the periodicity. Materials used in thin film technologies are usually characterized for, and with, an illumination at normal incidence. While nanometer to micron thick layers of material might not show any significant absorption, even minor residual absorptions might become problematic when aiming at long scale in plane propagation. This loss mechanism is extrinsic to the surface mode and might ideally be tailored by appropriate choice of the materials.
An ideal structure that would consist of lossless dielectric materials and perfect interfaces, i.e. free from any extrinsic loss mechanism, should only be limited by the authorized radiation through the substrate.

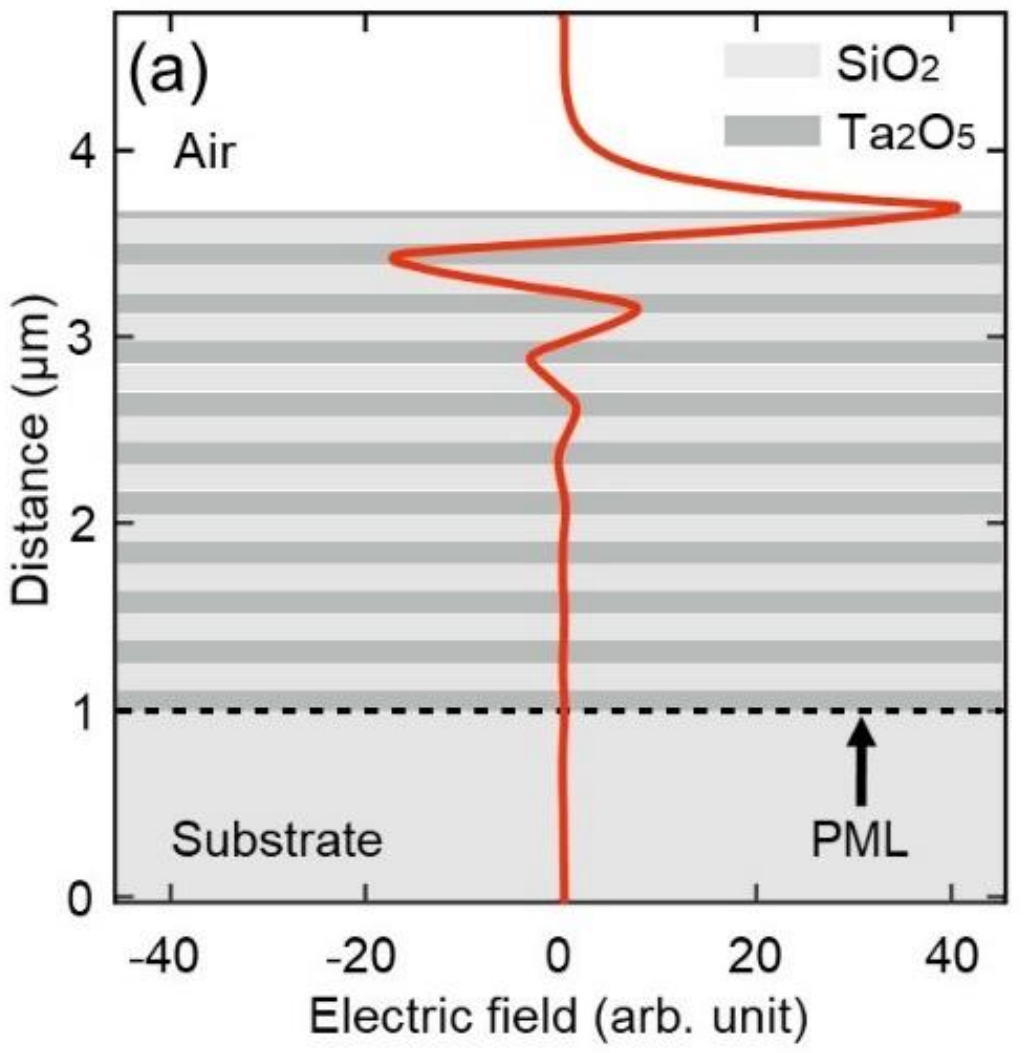

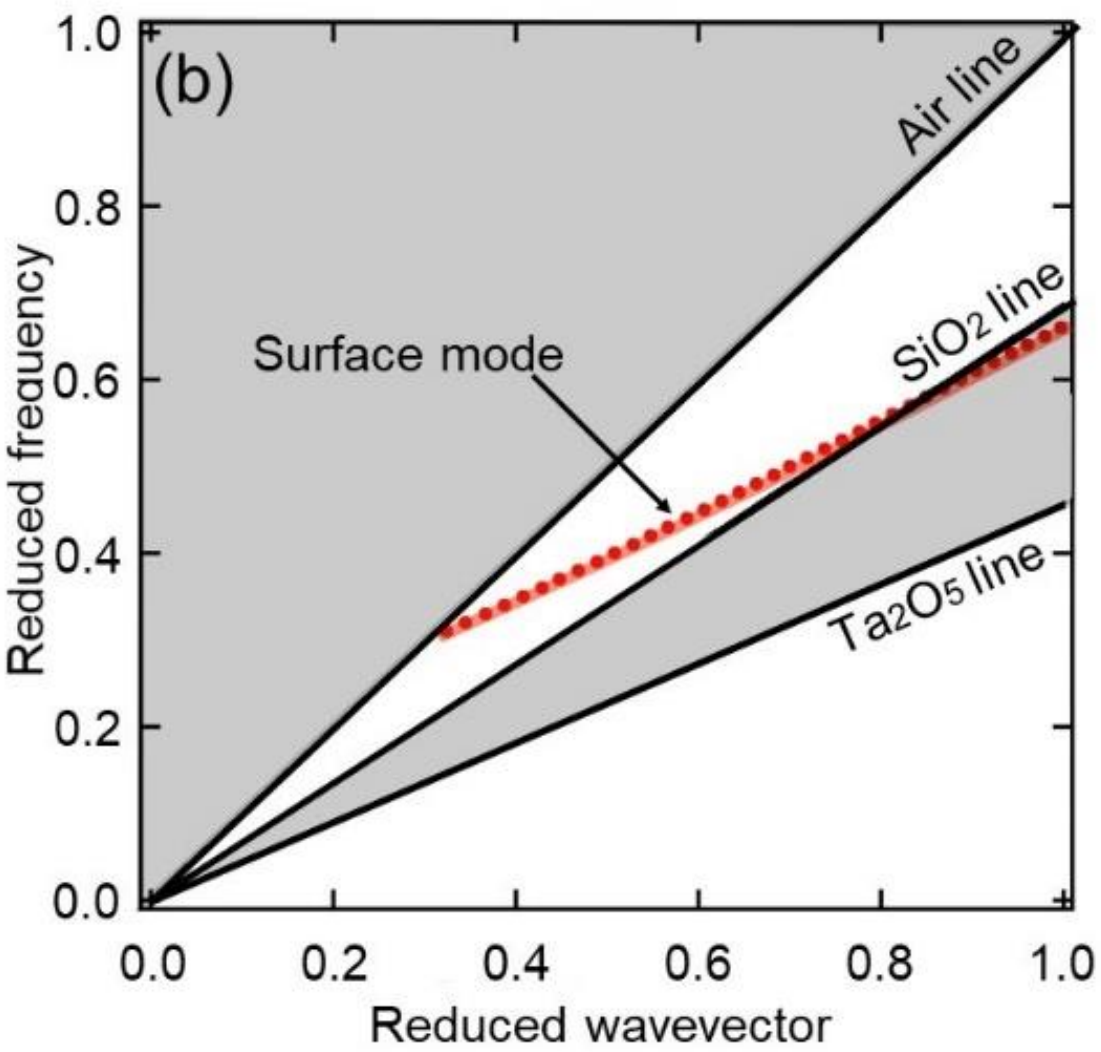

**Figure 1.** Optical surface waves on one-dimensional photonic crystals. a) Calculated distribution of the transverse electric field for a TE-polarized Bloch surface mode at a wavelength of 633 nm. The multilayer structure considered here consists of a one-dimensional photonic crystal with 10 lattice periods. Each period comprises of a 158 nm silicon dioxide ($SiO_2$) layer and a 107 nm tantalum pentoxide ($Ta_2O_5$) layer. A final 34 nm thick $Ta_2O_5$ layer is deposited on top of the structure. The structure is deposited on a glass substrate. b) Dispersion diagram of the surface mode presented in Figure 1a. The x and y axis are reported in reduced unit: reduced wavevector $k_{\mathrm{red}} = \frac{k\,\mathrm{a}}{2\pi}$ and reduced frequency $\omega_{\mathrm{red}} = \frac{\omega\,\mathrm{a}}{2\pi c}$ where a is the lattice period of the structure. The dispersion curve of the surface mode is shown as red dots. It extends from the air line on the low reduced frequency side to the $SiO_2$ line and beyond as the reduced frequency increases. Other modes sustained by the structure are not shown for the sake of clarity.

We use a 1D model to evaluate the impact of radiation losses. Leakage radiation can be elegantly simulated by using a perfectly matched layer (PML) boundary condition that acts as a perfect absorber at the interface between the surface mode and the mode continuum of the substrate. The complex propagation constant of the surface mode is calculated as the output of the model $k_{\mathrm{BSW}} = k_{\mathrm{real}} + \mathrm{i}\, k_{\mathrm{im}}$. The layers are modeled as perfect interfaces of lossless materials, and thus any technology-related loss is excluded. The model is expected to give an upper limit for the propagation losses of BSWs. Conventionally, the propagation length of a mode is calculated as the distance where the mode intensity decays by a factor of 1/e, in analogy to Beer-Lambert's law. Following this analogy, it is common to relate the absorption coefficient $\alpha$, to the imaginary part of the wavevector $\mathrm{k}_{im}$ [39]. The propagation length can then be written as: $L_{\mathrm{BSW}} = \frac{1}{\alpha} = \frac{1}{2\, k_{\mathrm{im}}}$.

**2.2. Design of long propagating optical surface waves.** The imaginary wavevector of a surface mode, for a given multilayer structure, can be fed into an optimization process. Structures supporting modes with minimal losses can thus be designed. The design routine relies on three main parameters, which are: (i) the lattice parameters of the periodic structure, (ii) the number of lattice periods, and (iii) the thickness of the top most layer. The lattice parameters are chosen based on the materials used for the layer deposition, and more specifically, based on their refractive indices. The optimization process first establishes rough lattice parameters depending on the desired operation wavelength. The thickness of the top most layer is then chosen such that a satisfying trade-off is achieved between (i) the level of confinement of the field at the interface of the multilayer and the top medium (air), (ii) the calculated propagation length, and (iii) the coupling angle required to couple to the surface mode through a prism. The structure is designed to operate around a central wavelength of 633 nm. The calculated coupling angle for this structure is 63.6° in a fused silica prism at 633 nm which corresponds to an angle of 72.5° in air using a right-angle prism. Figure 1 a and b correspond to the field distribution and dispersion diagram of a structure that is optimized for long propagation and large surface electric field simultaneously.

The propagation length of a one-dimensional photonic crystals having identical lattice parameters as the structure shown in Figure 1, but with a number of lattice periods ranging from 2 to 10 are computed (see Figure 2a)). Each computation takes only a few seconds using a personal computer. The plot is shown in semilogarithmic scale. An intuitive exponential decrease of the loss term $k_{im}$ can be observed as the number of lattice periods increases. The latter corresponds to a vanishing of the overlap between the surface state eigenmode and the radiation continuum. This effect translates into an exponential increase of the theoretical propagation length of the mode. Assuming thus that radiation into the substrate is the only limiting mechanism, it is found that a structure with 7 lattice periods should be enough to ensure a propagation length greater than 4 cm. Furthermore, it indicates that a technologically feasible structure consisting of 10 lattice periods should sustain the propagation of a surface mode for over 1 m.

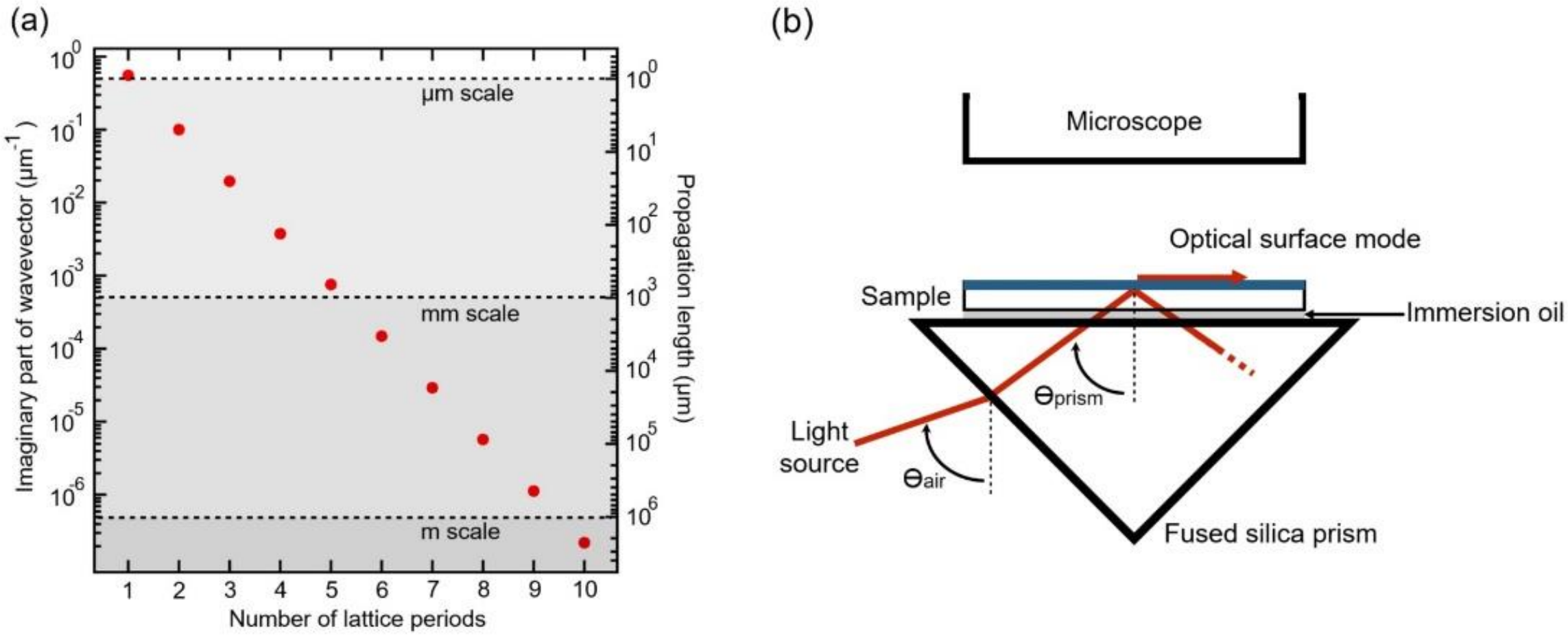

**Figure 2.** Evolution of the imaginary part of the wavevector $k_{im}$ as a function of the number of lattice periods, and schematic for experimental measurement. a) Imaginary part $k_{im}$ of the wavevector and associated propagation length of the Bloch surface mode as a function of the number of periods. Both quantities are plotted on a logarithmic scale. The simulation predicts an exponential increase of the propagation length when the number of periods is increased. This prediction is consistent with a decrease of the surface mode overlap with the radiation continuum, assuming that no other loss mechanism becomes predominant. b) Schematic of the experimental apparatus used to characterize the fabricated structure. Excitation of the surface mode is performed at a controlled angle through a fused silica prism and index matching oil. Propagation of the surface mode is confirmed and monitored using a specifically assembled microscope located above the investigated structure.

**2.3. Experimental demonstration of long propagating optical surface waves.** In order to experimentally validate our assumptions, a set of multilayer stacks with identical structural parameters but increasing number of lattice periods, from 4 to 7, is fabricated. An important part of this study relies on the capacity to isolate the losses induced by the predicted, intrinsic, radiation into the substrate from the two other extrinsic mechanisms. Ion beam sputtering technique is used to deposit the thin film multilayer structures and ensure high surface quality and low material absorption (see fabrication details in the Methods section).
The fabricated structures are characterized using a specifically assembled setup operating in a Kretschmann-type coupling configuration (see Figure 2b). As a first step, the angle at which a 633 nm radiation couples to the surface mode is measured. According to our eigenmode computations, coupling into the surface mode is expected to occur at an angle of 63.6°. Experimentally, excitation of the optical surface mode is observed for an incident angle of 63.9° ± 0.5°. This very good agreement with the numerical predictions validates the precision of the layer deposition process. Figure 3a shows a set of optical micrographs aggregated into a single image. Each micrograph corresponds to the optical image of the propagating surface wave acquired on each of the four samples fabricated. The image stack is assembled such that all the surface wave coupling points, where the excitation beam intersects the surface plane, are vertically aligned along the d = 0 line. The images are shown in false colors. The influence of the number of lattice periods can be clearly visualized on this figure. In particular, it is possible to observe that, starting from N = 6, the propagation of the surface wave extends over distances greater than 2 mm. In the case of the structure with 7 periods, the propagation extends well beyond the drawn 1 cm line.
Figure 3b shows the relative intensity decay profile of each sample. The latter is obtained by averaging the scattered intensity recorded in ten distinct positions of the sample to get rid of the effect of local scattering centers. The intensity variation is reported in dB, while the distance on the x-axis is reported in millimeters. Black dashed lines are added and serve as guide for the eyes. The propagation loss in dB/mm can be determined for each sample by directly measuring the decay per distance unit. The associated propagation length is reported in Figure 4. The strong reduction of the propagation losses when increasing the number of lattice period is clearly visible. The associated increase in propagation length predicted by our 1D eigenmode model is also confirmed qualitatively. A record propagation length of 1.4 cm is achieved for the structure consisting of 7 lattice periods. This figure surpasses by well over one order of magnitude the longest optical surface wave previously reported in the visible. It is also over 2 orders of magnitude greater than figures achieved with standard SPPs at equivalent optical frequencies. This result demonstrates that dielectric optical surface waves have the potential to transport optical information on a chip scale, while being noticeably more sensitive than guided waves to active or passive surface modifications. This last point offers room for development of more efficient and more compact sensors, detectors, or electro-optical modulators, for example.

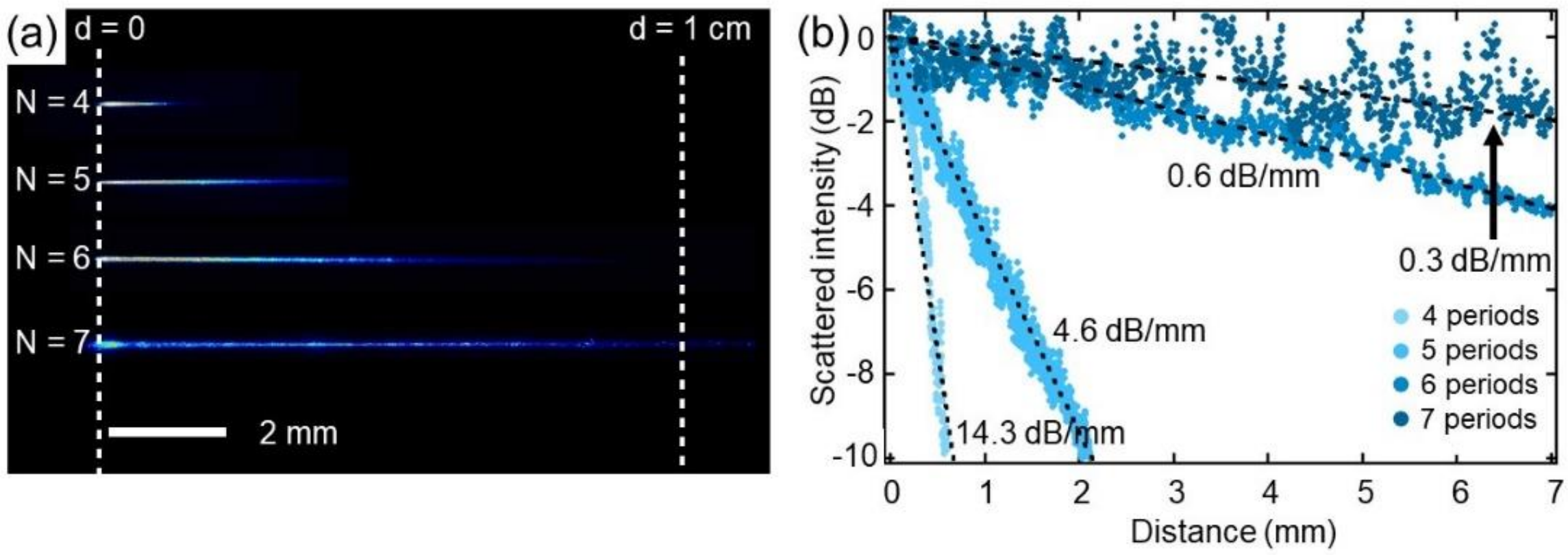

**Figure 3.** Optical surface waves characterization. (a) Images of the propagating Bloch surface waves (false colors). Images of the propagation recorded using the experimental setup for each of the four samples are stacked together and labelled according to the number of lattice periods in the structure. A white dashed line indicates the point where the surface mode is excited. A second dashed line indicates a distance of 1 cm from the injection point. One can observe that the structure with 7 lattice periods allows propagation of the surface mode beyond one centimeter. (b) Scattered intensity profile as a function of the distance from the injection point for each of the four fabricated samples. The scattered intensity is reported in dB so as to ease the reading of the propagation loss. A minimum of 0.3 dB/mm is achieved for the structure having 7 lattice periods.

As a separate validation, the propagation lengths of structures with a number of lattice periods ranging from 2 to 6 are computed using a rigorous time-domain solver (see Methods). Computation times significantly increase from a few minutes for the 2-period structure to a few days for 6 periods. Larger structures have not been calculated because of the very long computation time.

The calculated propagation lengths obtained using the two numerical methods are reported as a function of the number of lattice periods in Figure 4 along with the experimental data (red disks). Two main observations can be made. First, a fair agreement can be observed between the two numerical methods. There is as well a good agreement between the two numerical methods and the experimental data. This point validates our original assumption of radiation limited propagation and also our simple approach of introducing a perfectly matching layer acting as a perfect absorber in order to mimic the radiation losses in our 1D eigenmode solver-based model. As one could expect, the purely 1D eigenmode calculations provide slightly more optimistic predictions than the 3D time-domain calculations. Nevertheless, the use of our 1D approach over the 3D time-domain method is a significant gain in terms of modelling complexity and comes with a significant benefit: calculations times are reduced from days to seconds and, as a consequence, rapid iteration cycles and optimizations can be performed.

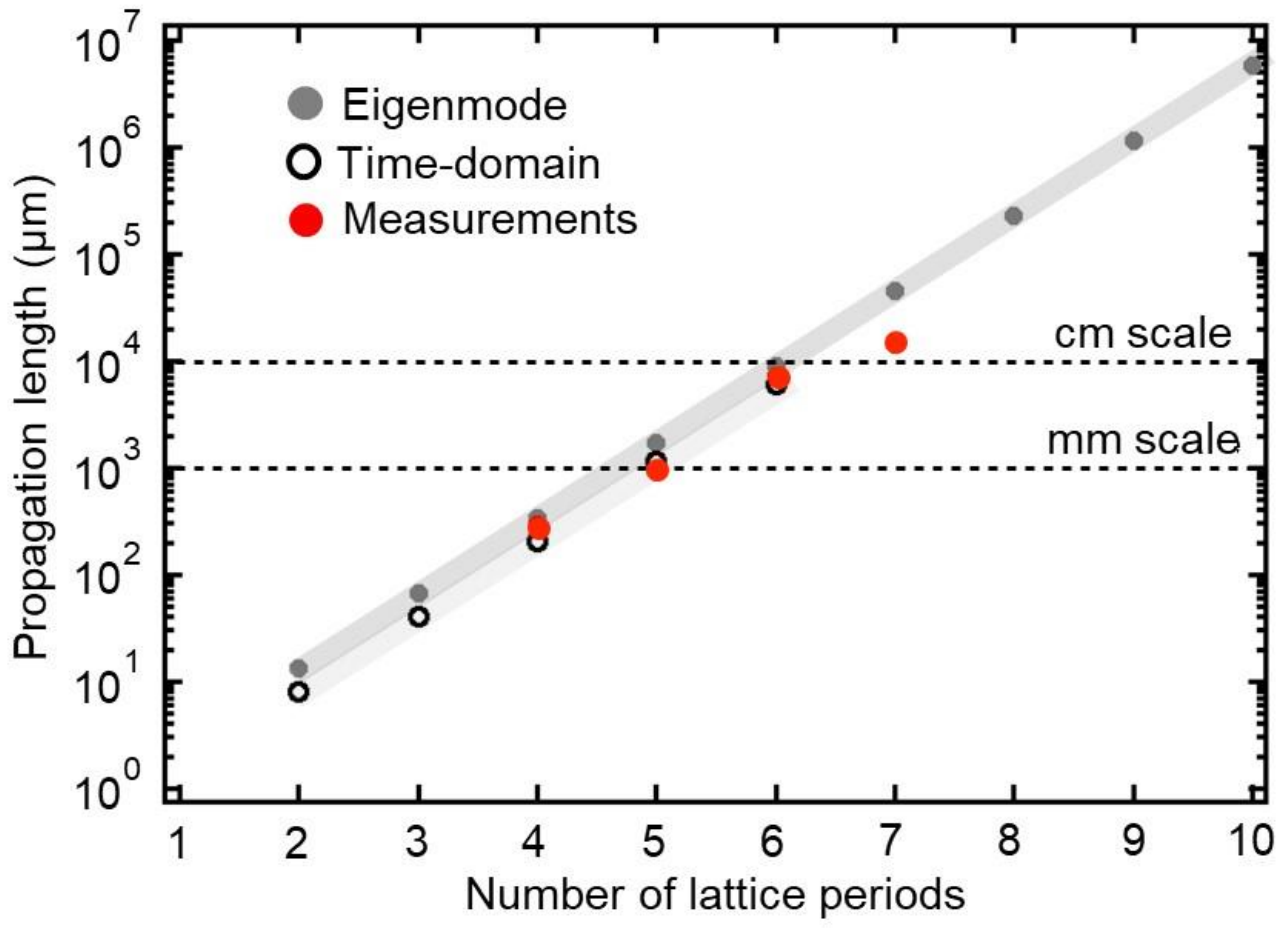

**Figure 4.** Comparison between the experimental and numerical propagation lengths. The propagation lengths are reported in log-scale as a function of the number of lattice periods in the multilayer structure. The numerical values are reported as gray disks for the eigenmode solver, and black circles for the time-domain solver. The experimental measurements are reported as red disks. The experimental and numerical data are in very good and quantitative agreement up to 6 lattice periods. After this point, a discrepancy seems to appear between the measured and numerical values, which might indicate the presence of another propagation limiting mechanism.

The experimental measurements precisely follow the numerical predictions obtained with the two independent methods, up to N = 6. The results shown in Figure 4 thus validate the prediction of exponential increase of the propagation length with the number of lattice periods (see Figure 2a), and our initial hypothesis that the major loss mechanism corresponds to the radiation of light through the multilayer. For the structure having 7 periods however, a discrepancy between the predicted (4.5 cm) and the measured (1.4 cm) propagation length appears. The nature of this discrepancy has not been identified at this point. It could possibly be attributed to another loss mechanism becoming dominant. Scattering, whether along the top surface or at layer interfaces, as well as residual material absorption, might be intuitively considered. For example, a set of early calculations based on the 1D model shows that material absorption levels of the order of $Im(n) = 10^{-5}$ , which are below our measurement capabilities, could be consistent with this observation. More importantly, this observation tends to indicate that any attempt to further improve the propagation length of BSWs in the visible might require an additional effort to further improve the surface quality and the residual material absorption. It might also be of interest to investigate the impact of scattering at the interfaces by replacing the standard fused silica wafers used in this experiment with superpolished substrates. The substrate roughness, before deposition, would then be reduced from the nanometer to Ångstrom levels thus potentially significantly the associated scattering contribution.

## 3. Discussion

Optical surface waves are challenging due to the strict conditions for their occurrence, but highly appealing because of the great overlap they provide with the top medium which is an advantage for light-matter applications. Here, we used a numerical optimization approach to design long propagating optical surface modes located between the air line and $SiO_2$ line. Taking an optimization approach was possible by distinguishing the intrinsic loss mechanism from the extrinsic ones, and by introducing a simple technique to numerically calculate the propagation length without requiring heavy computations. It opens new areas for future work where more advanced optimization approaches, i.e., inverse photonic design techniques can be employed to design Bloch surface waves platforms. Moreover, the high quality of the layers of the one-dimensional photonic crystal obtained by ion beam sputtering technique allowed us to experimentally demonstrate an exponential increase of the propagation length with the isolation of the surface mode, up to a certain point where other limiting mechanisms take over. Another interesting point can be raised from this study. The results presented here demonstrate that leakage radiation limits the propagation of a carefully designed and fabricated structure supporting a Bloch surface mode at (k,ω) coordinates located in between the air line and the substrate line. It might thus be of great interest to investigate the propagation properties of Bloch surface modes being operated below the substrate line, that is, where radiation of the surface mode into the substrate is simultaneously prohibited by multiple scattering and by total internal reflection. In this region, where $\omega_{\mathrm{red}} < \frac{k_{\mathrm{red}}}{n_{\mathrm{substrate}}}$ (here in between the $SiO_2$ line and the $Ta_2O_5$ line), the field is confined nearer to the surface and no radiation into the substrate is allowed. It is expected that a mode operated in this region (non-radiating regime) exhibits a propagation length equivalent to that of a mode operated in the radiating regime despite fewer lattice periods (e.g. 3 instead of 7). Such a structure would, on one hand, benefit from a simpler fabrication process and possibly from reduced number of fabrication-related defects and layer strain. On the other hand, the greater magnitude of the wavevector in this region prohibits the use of a Kretschmann-type excitation scheme, and thus imposes the use of another coupling scheme, e.g. a grating coupler. However, the full potential of such a platform may be harvested through the integration of a light source into the surface layer.

## 4. Conclusion

We have demonstrated, for the first time, the propagation of optical surface waves over centimeter-scales at visible wavelengths. We have focused our study on Bloch surface waves which are supported at the surface of a one dimensional photonic crystal. We report a 1.4 cm propagation length for a structure comprising of 7 lattice periods. This figure is well over one order of magnitude greater than the longest optical surface wave previously reported in the visible, and two orders of magnitude greater than standard surface plasmon polaritons. Our results open new perspectives for the use of optical surface waves for in-plane optical distribution and processing of information. In particular, it might motivate further work on the on-chip interfacing of Bloch surface states with other optical surface states.

## 5. Methods

**Thin-film multilayer deposition.** An ion beam sputtering machine (Navigator 1100, CEC GmbH) is used to produce the thin film multilayer structures and ensure high surface quality and low material absorption. Xenon is used as the sputtering gas. According to the designed structure, tantalum pentoxide ($Ta_2O_5$, n = 2.16 at λ = 633 nm) as high-refractive index and silicon dioxide ($SiO_2$, n = 1.49 at λ = 633 nm) as low-refractive index material are used. Ellipsometric combined with transmission measurements are performed on layers of the two materials in order to precisely assess the refractive indices and absorption coefficients of each material. The measurements indicate that no absorption is measurable on any of them, down to a level of Im(n) $=10^{-4}$ (imaginary part of the refractive index) which is the minimum detectable quantity. The oxides are formed by oxidation of metallic Ta and Si released from the sputtering targets with a deposition rate of about 0.1 nm/s. During the deposition, the vacuum pressure does not exceed $2 \cdot 10^{-3}$ mbar and the substrates are heated to 150 °C. The automated coating process is precisely controlled by broadband optical monitoring. The quality of the surface is assessed using an atomic force microscope (AFM). The average roughness of the surface is found to be 1 nm.

**Experiment.** A schematic of the excitation and monitoring part is shown on Figure 2b. Excitation of the surface modes is performed using a Kretschmann-type configuration based on a 45° fused silica prism and a goniometer. The light source consists of a supercontinuum source from NKT Photonics equipped with a custom acousto-optical tunable filter. The light source emission is centered around 633 nm with a 1.5 nm bandwidth. The polarization of the light is controlled using a polarizer and coupled into a polarization maintaining fiber. The polarized light is first collimated and then refocused using a 100 mm focal length lens through the prism onto the surface of the BSWs platform. The propagation length of the optical surface mode at the surface of the multilayer is measured with a custom microscope detection system. The latter collects the residual light that is scattered from the surface as the surface mode propagates. Imaging is performed using a high sensitivity scientific CMOS camera (Hamamatsu Orca Flash 4 V2). The measurement of the intensity profile, and the determination of the decay constant, is complicated by the presence of local variations in the scattered intensity induced by localized surface defects. To circumvent this issue, ten different acquisitions are performed, for each sample, in ten different locations of the surface. The ten images are then summed together which leads to a strong reduction of the noise. The resulting image is used to retrieve the decay constant.

**3D time-domain simulation.** To model the propagation lengths of the designed structures using a rigorous time-domain solver (CST Microwave Studio), a perfectly matching layer (PML) boundary condition is applied to the exterior boundaries. The multilayer is illuminated using a Gaussian beam at the designed coupling angle. The fields that radiate through the substrate reach the PML where they are absorbed. No material absorption or surface roughness effects is considered in this model to mimic the computations performed using our eigenmode-based model. A decay of the electromagnetic field intensity is observed along the propagation of the surface mode and, here again, the calculated propagation length corresponds only to the radiation loss through the multilayer. The rigorous time-domain modeling is greatly time-consuming, particularly for longer propagation lengths. For this reason, time-domain calculations have only been performed for multilayers with up to 6 layer pairs.

## Acknowledgements

The authors acknowledge Raphaël Monnard for the AFM measurements of the samples. The work of Nicolas Descharmes and Raphaël Barbey is supported by the Gebert Rüf Stiftung, project number GRS 019/16.